\documentclass[aps,prl,twocolumn]{revtex4}
\usepackage{epsfig}

\begin{document}
\headsep 2.5cm
\title{New Real space method for calculation of physical properties of a disordered system.}

\author{Rostam Moradian}
\email{rmoradian@razi.ac.ir}
\affiliation{$^{1}$Physics Department, Faculty of Science, Razi
University, Kermanshah, Iran\\
$^{2}$Computational Physical Science Research Laboratory, Department of Nano-Science, Institute for Studies in Theoretical Physics and Mathematics (IPM)
            ,P.O.Box 19395-1795, Tehran, Iran.}

\date{\today}

\begin{abstract}
We introduce a new real space super cell approximation method for
treating the electronic states of disordered systems. This method
is general and allows both  randomness in the on-site energies and
in the hopping integrals. In the special case of randomness in the
on-site energies only, this method is equivalent to the Non Local
Coherent Potential Approximation (NLCPA) derived previously.
 \end{abstract}

\maketitle

For the past three decades the Coherent Potential Approximation
(CPA)  was the most widely used self-consistent mean field treatment of
disordered systems\cite{Soven:67}. As a single site approximation,
the CPA Green's functions are averaged over the randomness on a
single site, and so could not handle the effects of correlations
between the random potentials on neighbouring sites. In order to
include such inter site effects extensions of CPA were developed
using the locator
formalism\cite{Blackman:71,Esterling:75,Gonis:77}. However,
recently the Dynamical Cluster Approximation (DCA) has been
introduced   for interacting systems\cite{Hetler:98}. A version of
DCA has also been used to treat disordered systems
\cite{Jarrell:01}. This method extends CPA by including a weakly
wave vector dependent self energy $\Sigma(k;\omega)$. The DCA was
originally derived on basis of restricting the role of momentum
conservation in calculating the self-energy, by applying a coarse
graining procedure to the First Brillouin Zone (FBZ). Momentum
conservation is applied only to the coarse grained wave vector
${\bf K}_{n}$ in the FBZ and in the corresponding Laue function.
In this method a key computational step makes use of a Fourier
like transformation of coarse grained wave vectors ${\bf K}_{n}$ to a
real-space set of coordinates, ${\bf R}_{n}$. These coordinates
were interpreted as the lattice sites of a real cluster of a real
lattice, however this connection was not established directly.

By using effective medium theory in real-space we developed a
non-local CPA  (NLCPA) method. In this method a cluster of
impurities is  embedded in an effective medium, in such a way that
all cluster sites are equivalent and lattice periodicity is
maintained for the disorder averaged Green functions.  For a one
dimensional binary alloy numerically we showed that a periodic
super cell density of states is similar to the NLCPA density of
states\cite{Moradian2:00}. We also applied NLCPA as a real space
super cell approximation to explain the resonance peak that
appears in the density of states of substituted $Zn$ impurity in
$CuO_{2}$ plane of
$Bi_{2}Sr_{2}Ca_{2}Cu_{3}O_{8+\delta}$\cite{Moradian:02}.
Subsequently NLCPA was also extended to the case of the
first-principles Korringa-Kohn-Rostoker (KKR) CPA model for the
electronic structure of disordered systems.\cite{Rowlands:03}

In this paper we show that the super cell self energy periodicity
in real space leads naturally to coarse graining in k-space,
providing us with an alternative and general derivation of the
NLCPA method. This establishes directly the connection between the
real-space cluster and the coarse graining in k-space, showing
clearly for the first time that both DCA and NLCPA are equivalent
to a super cell approximation in real space.

We start our investigation on a general tight binding model for a
non-interacting alloy system,
\begin{equation}
H=-\sum_{ij}t_{ij}c^{\dagger}_{i\sigma}c_{j\sigma}+\sum_{i\sigma} (\varepsilon_{i}-\mu)
 \hat{n}_{i\sigma} .
\label{eq:Hamiltonian}
\end{equation}
where $t_{ij}$ are the hopping integrals which may be random,
$\mu$ is the chemical potential and $\varepsilon_{i}$ is the
random on site energy which takes values  $-\delta/2$ with
probability $1-c$ for host sites, and $\delta/2$ with probability
$c$ for impurity sites.

 The equation of motion for electrons with the above Hamiltonian, Eq.\ref{eq:Hamiltonian}, for any
 general impurity configuration $\{\varepsilon_{i}\}$ in terms of  the Green function is,
\begin{equation}
\sum_{l} \left(
       \begin{array}{c}
(\omega-\varepsilon_{i}+\mu)\delta_{il}+t_{il}\end{array}\right){ G}(l,j;\omega)=\delta_{ij}
\label{eq:equation of motion}
\end{equation}
 We expand the full random Green function $G(i,j;\omega)$ in Eq.\ref{eq:equation of motion},
 in terms of the clean system Green function ${G}^{0}(i,j;\omega)$,\cite{Elliott:74}
\begin{equation}
 G(i,j;\omega)={ G}^{0}(i,j;\omega)+\sum_{ll^{'}}{ G}^{0}(i,l;\omega)
{V}_{ll^{'}}{G}(l^{'},j;\omega)
\label{eq:expanding interms of random onsite potential}
\end{equation}
where the random potential matrix ${ V}_{ll^{'}}$ is
\begin{equation}
{V}_{ll^{'}}=\begin{array}{c}
\varepsilon_{l}\delta_{ll^{'}}+\delta t_{ll^{'}}
\end{array},
\label{eq:random onsite potential}
\end{equation}
$\delta t_{ll^{'}}=t_{ll^{'}}-t^{0}_{ll^{'}}$ is the difference between hopping integral in random and clean
 system and ${ G}^{0}(i,j;\omega)$ is
\begin{equation}
{ G}^{0}(i,j;\omega)=\frac{1}{N}\sum_{\bf k}e^{\imath{\bf k}.{\bf r}_{ij}}
\left(\begin{array}{c}
\omega-\epsilon_{\bf k}+\mu
\end{array}\right)^{-1} .
\label{eq:clean}
\end{equation}
where $\epsilon_{\bf k}=-\frac{1}{N}\sum_{ij}t^{0}_{ij}e^{{\bf
k}.{\bf r}_{ij}}$ is the band structure for the clean system.

The Dyson equation corresponding to Eq.\ref{eq:expanding interms of random onsite potential} is
\begin{equation}
 \bar{G}(i,j;\omega)={ G}^{0}(i,j;\omega)+\sum_{ll^{'}}{ G}^{0}(i,l;\omega)
\Sigma(l,l^{'};\omega){\bar G}(l^{'},j;\omega)
\label{eq:Dyson equation}
\end{equation}
where,
\begin{equation}
\sum_{l^{'}}\langle {V}_{ll^{'}}G(l^{'},j;\omega)\rangle = \sum_{l^{'}}\Sigma(l,l^{'};\omega){\bar G}(l^{'},j;\omega).
\label{eq:self energy definition}
\end{equation}
The disorder average Green function, Eq.\ref{eq:Dyson equation},
can be written,
 \begin{equation}
{ \bar G}(i,j;\omega)=\frac{1}{N}\sum_{\bf k}e^{\imath{\bf k}.{\bf r}_{ij}}
\left(\begin{array}{c}
\omega-\epsilon_{\bf k}+\mu-\Sigma({\bf k};\omega)
\end{array}\right)^{-1} .
\label{eq:exact average green function}
\end{equation}
where
 \begin{equation}
\Sigma({\bf k};\omega)=\frac{1}{N}\sum_{i,j}e^{-\imath{\bf k}.{\bf r}_{ij}}
\Sigma(i,j;\omega) .
\label{eq:self energy fourier transform}
\end{equation}
is the  Fourier transform of the self-energy.

Fourier transformations of this system of equations obey the
orthogonality condition,
\begin{equation}
\frac{1}{N}\sum_{\bf k}e^{-\imath{\bf k}.{\bf r}_{ij}} =\delta_{ij}
\label{eq:wave vector orthogonality}
\end{equation}
where indices $i$ and $j$ run over all $N$ lattice sites in the crystal.
Because the exact solution of
Eqs.\ref{eq:expanding interms of random onsite potential} and
\ref{eq:Dyson equation} is impossible, they are usually solved in
different single site approximations such as the Born
approximation, the T-matrix approximation and the
CPA.\cite{Moradian2:00} It is well known that in all these
approximations inter site impurity effects has been neglected.

\begin{figure}
\centerline{\epsfig{file=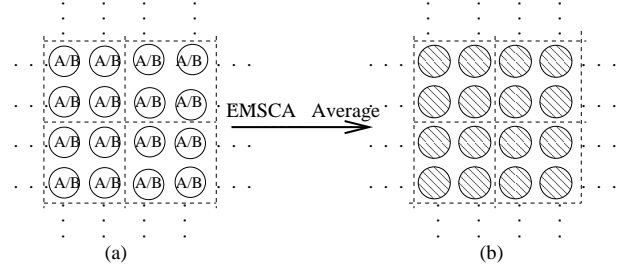,width=8.0cm,angle=0}}
\caption{(a): A binary alloy divided into super cells containing
four sites. (b) The approximation of  impurity averaging over the
super cells. By neglecting the multiple scattering between
impurities in different super cells, the self energies in
different super cells obey the periodicity law given in
Eq.\ref{eq:self periodicity}.
\label{Fig:randomlatticesupercell-average}}
\end{figure}

 Here we introduce a new approximation, and we show that, for the case of $\delta t_{ll^{'}}=0$, it is
 equivalent to the NLCPA \cite{Moradian:02}. The derivation of the approximation is as follows. Consider a random
 alloy system that has been divided to  $N_c$ super cells.  The super cell Green function ${G}_{sc}(I,J;\omega )$ can be expand as
\begin{eqnarray}
{G}_{sc}(I,J;\omega )&=&
 {G}^{0}(I,J;\omega)\nonumber\\&+&\sum_{l,l^{\prime }}{G}^{0}(I,l;\omega)V _{ll'}{G}_{sc}(l^{\prime},J; \omega).\nonumber\\
\label{eq:random impurity super cell Green function}
\end{eqnarray}
where I and J refer to sites in the same super cell  and
$l$,$l^{'}$ indices refer to the whole lattice. The average Green
function corresponding to Eq.\ref{eq:random impurity super cell
Green function} is,
\begin{eqnarray}
{\bar G}_{sc}(I,J;\omega )&=&
 {G}^{0}(I,J;\omega)\nonumber\\&+&\sum_{l,l^{\prime }}{G}^{0}(I,l;\omega)\Sigma_{sc}(l,l^{\prime };\omega){\bar G}_{sc}(l^{\prime},J; \omega).\nonumber\\
\label{eq:average super cell Green function}
\end{eqnarray}
Now we apply an approximation that is based on two assumptions:
first {\em {neglect the hopping integrals deviation,$\delta
t_{ll^{'}}$, between impurity sites  of different super cells,
that is, $\delta t_{ll^{'}}=0$ when $l$ and $l^{'}$ are in
different super cells }} and second {\em{neglect multiple
scattering between impurities in different super cells}}.
Therefore we have
\begin{equation}
\Sigma_{sc}(i,j;\omega) =0,\;\; if\;\; i\; and \;j\; \notin \;same\; super\;\; cell.
\label{eq:different sites self energy}
\end{equation}
 Fig.\ref{Fig:randomlatticesupercell-average} shows a binary alloy system divided into periodic super cells containing
 four impurity sites. By neglecting both the hopping integrals deviations $\delta t_{ll^{'}}$ and also the correlations
 between different super cells sites, the self energy in each super cell is independent of the other super cells and is periodic
 with respect to super cell translation vectors.
\begin{figure}
\centerline{\epsfig{file=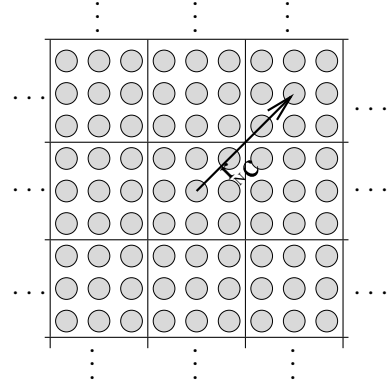,width=5.0cm,angle=0}} \caption{A
two dimensional lattice divided into super cells of nine sites.
The arrow indicates  the super cell periodicity vector ${\bf
r}_{N_c}=3{\bf a}_{1}+3{\bf a}_{2}$, where $N_{c1}=3$ and $N_{c2}=3$. \label{Fig:rnc}}
\end{figure}
These conditions imply that the disorder averaged system self
energy has the following super cell periodicity,
 \begin{equation}
\Sigma_{sc}({\bf r}_{IJ}+{\bf r}_{N_c};\omega)=\Sigma_{sc}({\bf
r}_{IJ};\omega)  \label{eq:self periodicity}
\end{equation}
where $I$ and $J$ refer to the sites in the same super cell, and
the super cell translation vector ${\bf r}_{ N_c}$ is
 \begin{equation}
 {\bf r}_{ N_c}=\sum_{i=1}^{3}\;m_{i}\;{N_c}_{i}\;{\bf a}_{i},
\label{eq:supercell}
\end{equation}
 where $N_c={N_c}_{1}\;{N_c}_{2}\;{N_c}_{3}$ is the number of lattice sites in a 3-dimensional super cell,
 ${\bf a}_{i}$ are primitive vectors of real lattice, and ${N_c}_{i}$ is the number of sites in the super cell in the $i$th direction
 and $m_{i}$ are integers. Fig.\ref{Fig:rnc} shows ${\bf r}_{N_{c}}$ for a two dimensional lattice with periodicity over
 nine sites in each super cell.  Note that when $N_c\rightarrow\infty$ all super cells
 will coincide and are equal to the full real lattice.

\begin{figure}
\centerline{\epsfig{file=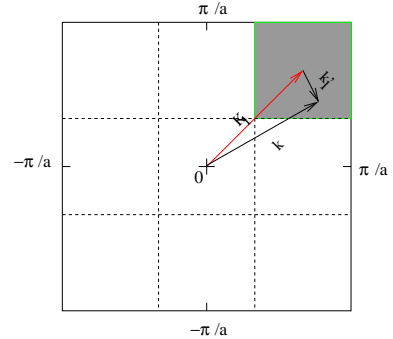,
width=5.0cm,angle=0}} \caption{Relation between ${\bf k}$, ${\bf
K}_{n}$ and ${\bf k'}_{n}$ in the First Brillouin Zone (FBZ) for a
two dimensional system with a 9 site super cell. As given by
Eqs.~\ref{eq:supercell wave vector} and \ref{eq:k-relation} there
are 9 different ${\bf K}_{n}$ in the FBZ, and so the  FBZ is
divided in to 9 equal areas,  while each ${\bf K}_{n}$ is located
at the center of one of these areas. \label{Fig:pupercellNc9-Bz}}
\end{figure}

 Fourier transformation of Eq.\ref{eq:self periodicity} implies that we should restrict the self energy wave vectors ${\bf k}$ to
 those obeying
\begin{equation}
e^{\imath{\bf k}.{\bf r}_{\tiny N_c}} =1, \label{eq:wave vector
definition}.
\end{equation}
These are the set
\begin{equation}
{\bf K}_{n}=\sum_{i=1}^{3} \frac{l_{i}}{{N_c}_{i}} {\bf b_{i}}.
\label{eq:supercell wave vector}
\end{equation}
where ${\bf b_{i}}$ are the reciprocal lattice primitive vectors
and $l_i$ are integers. Notice that number of the ${\bf K}_{n}$ in
the first Brillouin zone is equal to number of the lattice sites,
$N_c$, in a super cell.
 Therefore the Fourier transformation of the self energy is
\begin{equation}
\Sigma({\bf K}_{n};\omega)=\frac{1}{N}\sum_{i,j}e^{-\imath{\bf
K}_{n}.{\bf r}_{ij}}\Sigma_{sc}(i,j;\omega). \label{eq:k-self
energy}
\end{equation}
By inserting Eq.\ref{eq:different sites self energy} into
Eq.\ref{eq:k-self energy} we find that
\begin{equation}
\Sigma({\bf K}_{n};\omega)=\frac{1}{N_c}\sum_{I,J}e^{-\imath{\bf
K}_{n}.{\bf r}_{IJ}}\Sigma_{sc}(I,J;\omega) \label{eq:k-self
energy supercell}
\end{equation}
 where $I$ and $J$ refer to the sites in the same super cell.
 By converting the Eq.\ref{eq:k-self energy supercell} to  real space we find
\begin{equation}
\Sigma_{sc}(I,J;\omega)=\frac{1}{N_c}\sum_{{\bf K}_{n}}e^{-{\bf
K}_{n}.{\bf r}_{IJ}}\Sigma({\bf K}_{n};\omega) \label{eq:real
space self energy supercell}
\end{equation}
where the orthogonality condition in the super-cell is,
\begin{equation}
\frac{1}{N_{c}}\sum_{{\bf K}_{n}}e^{-\imath{\bf K}_{n}.{\bf
r}_{IJ}} =\delta_{IJ} \label{eq:supercell orthogonality}
\end{equation}
Because the set of $\{{\bf K}_{n}\}$ vectors divides the volume of
the first Brillouin zone to $Nc$ equal patches, we therefore
identify each patch, $n$, by the corresponding wave vector ${\bf
K}_{n}$ that is located at its center. Therefore we define the
relation between ${\bf K}_{n}$ and ${\bf k}$ inside each patch as
follows
\begin{equation}
{\bf k}={\bf K}_{n}+{\bf k}^{'}_{n}
\label{eq:k-relation}
\end{equation}
where ${\bf k}^{'}_{n}$ are the wave vectors inside of $n$th patch
with respect to the center of the patch.
Fig.\ref{Fig:pupercellNc9-Bz} illustrates this relationship,
Eq.\ref{eq:k-relation}, for one of the patches in a two
dimensional system for $N_c=9$.
By comparing Eqs.~\ref{eq:supercell orthogonality} and 
\ref{eq:wave vector orthogonality} we find that the super cell approximation is
equivalent to the replacement
\begin{equation}
e^{\imath {\bf k}^{'}_{n}.{\bf r}_{IJ}} \approx 1.
\label{eq:super cell approximation}
\end{equation}
 Considering of the  Eqs.~\ref{eq:super cell approximation}, \ref{eq:supercell orthogonality}, \ref{eq:k-self energy supercell} and Eq.~\ref{eq:average super cell Green function} we find that the super cell approximation Green function ${\bar G}_{sp}(I,J;\omega)$ is
 \begin{equation}
{\bar G}_{sc}(I,J;\omega)=\frac{1}{N_c}\sum_{{\bf K}_{n}}e^{\imath{\bf K}_{n}.{\bf r}_{ij}}
{\bar G}({\bf K}_{n};\omega)
\label{eq:supercell average green function}
\end{equation}
where
 \begin{equation}
{\bar G}({\bf K}_{n};\omega)=\frac{N}{N_c} \sum_{{\bf
k}^{'}_{n}}\left(\begin{array}{c} \omega-\epsilon_{{\bf
K}_{n}+{\bf k}^{'}_{n}}+\mu-\Sigma({\bf K}_{n};\omega)
\end{array}\right)^{-1} .
\label{eq:k-supercell average green function}
\end{equation}
In short Eq.~\ref{eq:k-self energy} and 
\ref{eq:super cell approximation} are the super cell approximation conditions. In the
limit of $N_c \rightarrow\infty$, ${\bf k}^{'}_{n} \rightarrow 0$
and also ${\bf K}_{n} \rightarrow {\bf k}$. Therefore
Eqs.~\ref{eq:supercell orthogonality} and
 \ref{eq:k-self energy supercell} respectively convert to Eqs.~\ref{eq:wave vector orthogonality} 
 and \ref{eq:self energy fourier transform} in the
exact average system. Also for the exact average system in the
limits of super cell periodicity with $N=mN_c$, with m an integer,
in the case where correlation between super cells is neglected,
the exact average self energy with these conditions  equals the
super cell periodic self energy, that is
\begin{equation}
\Sigma(I,J;\omega)=\Sigma_{sc}(I,J;\omega).
 \label{eq:equality of self energies}
\end{equation}
By  use of
 Eq.~\ref{eq:super cell approximation} the Fourier
transformation of Eq.~\ref{eq:equality of self energies} can be
carried out by reordering the summation over {\bf k} as $\sum_{\bf
k}=\sum_{\bf K_{n}}\sum_{{\bf k}^{'}_{n}}$ and using the fact that
$\sum_{{\bf k}^{'}_{n}} 1=\frac{N}{Nc}$. We find that
\begin{equation} \Sigma({\bf
K}_{n};\omega)=\frac{1}{N_c}\sum_{{\bf k}^{'}_{n}} \Sigma({\bf
K}_{n}+{\bf k}^{'}_{n};\omega)
 \label{eq:k-equality of self energies}
\end{equation}
and Eqs.~\ref{eq:wave vector orthogonality}, \ref{eq:self energy fourier transform} and 
\ref{eq:exact average green function}
will become equal to
 Eqs.~\ref{eq:supercell orthogonality}, \ref{eq:k-equality of self energies} and
\ref{eq:supercell average green function} respectively. This means
that the super cell calculation in an effective medium is an
approximation for the exact disordered system that includes only
the correlation between impurities within a single super cell.

\begin{figure}
\centerline{\epsfig{file=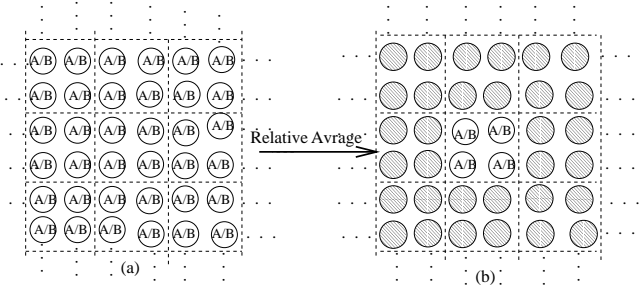, width=8.50cm,angle=0}}
\caption{A four site impurity super cell in an effective medium of mean-field super cells.
\label{Fig:clusterimpurity-average}}
\end{figure}

On the other hand, by eliminating the clean Green function matrix
${\bf G}_{0}$ between
Eqs.~\ref{eq:expanding interms of random onsite potential} and \ref{eq:Dyson equation} the random matrix
Green function ${\bf G}$ can be expressed in terms of the average
Green function   $\bar{\bf G}$ as
 \begin{equation}
  {\bf G}=\bar{\bf G}+\bar{\bf G}({\bf V}-{\bf \Sigma}){\bf G}.
\label{eq:random matrix Green}
\end{equation}
By applying the
super cell approximation with 
Eq.~\ref{eq:different sites self energy}, and  taking the average over impurities in all super cells
except one, Eq.~\ref{eq:random matrix Green} reduces to the
following equation for the remaining super cell,
 \begin{equation}
{\bf G}^{imp}_{sc}=\bar{\bf G}_{sc}+\bar{\bf G}_{sc}({\bf
V}_{sc}-{\bf \Sigma}_{sc}){\bf G}^{imp}_{sc}.
 \label{eq:super cell random matrix Green}
\end{equation} where all the matrices in
Eq.~\ref{eq:super cell random matrix Green} are of dimension $N_{c}\times N_{c}$. 
By rearranging Eq.\ref{eq:super cell random matrix Green}
we have 
\begin{equation}
 {{\bf G}^{imp}_{sc}}^{-1}+{\bf
V}_{sc}={\bar{\bf G}_{sc}}^{-1}+{\bf \Sigma}_{sc}.
\label{eq:arange matrix Green} 
\end{equation} 
Defining each side
of Eq.\ref{eq:arange matrix Green} as a cavity super cell Green
function ${\hat\mathcal{ G}}^{-1}$ we have 
\begin{equation} {{\bf
G}^{imp}_{sc}}^{-1}+{\bf V}_{sc}={\hat\mathcal{ G}}^{-1}.
\label{eq:super cell matrix Green} 
\end{equation}
 and
\begin{equation}
 {\bar{\bf G}_{sc}}^{-1}+{\bf
\Sigma}_{sc}={\hat\mathcal{ G}}^{-1}.
 \label{eq:super cell Dyson eq}
 \end{equation}
 Fig.~\ref{Fig:clusterimpurity-average} shows an example four site super cell approximation in which the average has been taken over all super cells except one {\em impurity super cell}. The matrix elements of Eq.~\ref{eq:super cell matrix Green} are
\begin{eqnarray}
{G}^{imp}_{sc}(I,J;\omega )&=&
{\mathcal {G}}(I,J;\omega)\nonumber\\&+&\sum_{L,L^{\prime }}{\mathcal{G}}(I,L;\omega)V _{LL'}{G}^{imp}_{sc}(L^{\prime},J; \omega).\nonumber\\
\label{eq:random impurity cluster Green function}
\end{eqnarray}
where the cavity Green function ${\mathcal{G}}(I,J;\omega)$ obeys
 \begin{equation}
{\mathcal{G}}(I,J;\omega)=\frac{1}{Nc}\sum_{{\bf K}_{n}}e^{\imath{\bf K}_{n}.{\bf r}_{IJ}}
{\mathcal{G}}({\bf K}_{n};\omega).
\label{eq:cavity supercell green function}
\end{equation}
Similarly, the matrix elements corresponding to Eq.~\ref{eq:super cell Dyson eq} are,
\begin{eqnarray}
\bar{G}_{sc}(I,J;\omega )&=&
{\mathcal {G}}(I,J;\omega)\nonumber\\&+&\sum_{L,L^{\prime }}{\mathcal{G}}(I,L;\omega)
\Sigma_{sc}(L,L^{\prime }; \omega)\bar{G}_{sc}(L^{\prime},J; \omega).\nonumber\\
\label{eq:coherent cluster Green function}
\end{eqnarray}
where the average of the impurity Green function ${G}^{imp}_{sc}(I,J;\omega )$ over all impurity configurations $\left\{ V_{IJ}\right\}$ in the super cell is
\begin{equation}
\langle{G}^{imp}_{sc}(I,J;\omega )\rangle={\bar G}_{sc}(I,J;\omega).
\label{eq:equality of green functions}
\end{equation}
Eqs.~\ref{eq:k-supercell average green function}, \ref{eq:k-equality of self energies}, \ref{eq:coherent cluster Green function}, \ref{eq:equality of green functions}  constitute a closed set to be solved self consistently.

The above system of equations can be implemented numerically by the following algorithm:\\
1- make a guess for $\Sigma({\bf K}_{n};\omega)$, usually zero.\\
2- Calculate ${\bar G}({\bf K}_{n};\omega)$ from Eq.~\ref{eq:k-supercell average green function},\\
3- Use Eq.~\ref{eq:coherent cluster Green function} to calculate the Fourier transform of the cavity Green function\\
\begin{equation}
{\mathcal{G}}({\bf K}_{n};\omega)=({\bar G}^{-1}({\bf K}_{n};\omega)+\Sigma_{sp}({\bf K}_{n};\omega))^{-1}
\label{eq:k-cavity green functions}
\end{equation}
4- Calculate the impurity super cell Green function ${G}^{imp}_{sp}(I,J;\omega )$ from 
Eq.~\ref{eq:random impurity cluster Green function}.\\
5- Calculate the average Green function from Eq.~\ref{eq:equality of green functions} and Fourier transform it to ${\bf K}_{n}$ space by using Eq.~\ref{eq:supercell orthogonality}.\\
6- Using Eq.~\ref{eq:k-cavity green functions} calculate the new self energies $\Sigma_{sc}({\bf K}_{n};\omega)$ and go back to step 1 and repeat whole process until convergence has been obtained to a desired accuracy.

\begin{figure}
\centerline{\epsfig{file= 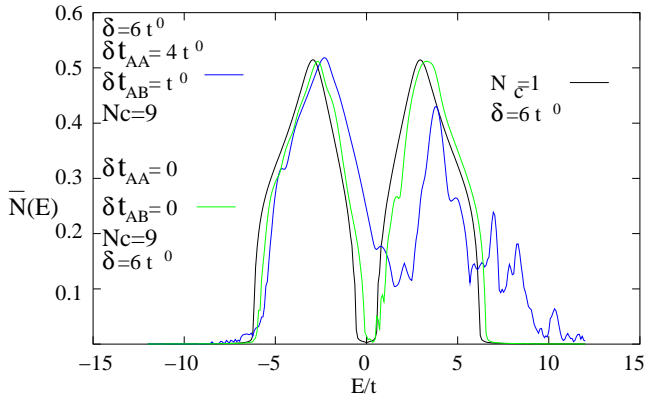 , width=8.50cm,angle=0}}
\caption{The density of states of a two dimensional square lattice of a binary alloy $A_{c}B_{1-c}$ at half band filling, ${\bar n}=1$, for $c=0.5$ and $\delta=6t^{0}$ when $N_{c}=1$ (CPA) and $N_{c=9}$. We compared  $N_{c}=1$ DOS and of two cases of $N_c=9$, first $\delta t_{AA}=0$, $\delta t_{AB}=0$ and second  $\delta t_{AA}=4t^{0}$, $\delta t_{AB}=t^{0}$. For both non hopping integral randomness cases the band splitting happened ( metal insulator phase transition take placed), while for the other not (still is a metal).  
\label{Fig:nor2dimcomtijNc9cpan1 }}
\end{figure}

As an application of this method we calculate the density of states of a two dimensional square lattice at half band filling, ${\bar n}=1 $ and $\delta=6t^{0}$ in which $N_c=1$ (CPA) and $N_{c}=9$. For the $N_c=9$ case, we also show the effects of introducing random hopping parameters, by considering the two cases, $\delta t_{AA}=0$, $\delta t_{AB}=0$ and $\delta t_{AA}=4 t^{0}$, $\delta t_{AB}=t^{0}$, where
\begin{equation}
\delta t_{\langle ij\rangle}=t_{AA}-t^{0}
\label{eq:deltatijAA}
\end{equation}
which $i$ and $j$ are nearest neighbour sites and both are A type, and we also defined, 
\begin{equation}
\delta t_{\langle ij\rangle}=t_{AB}-t^{0}
\label{eq:deltatijAB}
\end{equation}
where at the nearest neighbours sites $i$ and $j$ one is A and another is B type atom. Fig.\ref{Fig:nor2dimcomtijNc9cpan1 } shows that for the $Nc=1$ (CPA) and the case of $Nc=9$ without random hopping, in which  $\delta=6t^{0}$, we are at the band splitting regime\cite{ziman:79, Moradian:01}. In this case at half band filling a metal-insulator phase transition is taken place, in spite of different gap sizes. While for the case of $Nc=9$ with random hopping $\delta t_{AA}=4 t^{0}$, $\delta t_{AB}=t^{0}$, the band splitting is not happened, thus the system is a metal. Therefore due to including randomness in both the on-site energies and also the hopping integrals, EMSCA technique can be provide a more realistic results. 

Finally, we summarize our discussion with the following conclusions.  We have introduced a new real space effective medium super cell approximation (EMSCA) for random systems. For the case $N_{c}=1$ this recovers the single site CPA formalism.  However for larger values of $N_c$ the
method extends CPA by including both the effects of multiple scattering,and by allowing randomness in the hopping integrals. Furthermore, in the limit of $N_{c}\rightarrow\infty$ the method is exact. The EMSCA in the special case where hopping randomness is neglected, $\delta t=0$, leads to an alternative derivation of the NLCPA technique for disordered systems. This derivation completely establishes the NLCPA as a valid and useful extension of the old and  popular CPA method, which  incorporates the effects of
inter-site correlations.

\acknowledgements I would like to thanks James F. Annett and B. L. Gyorffy for helpful discussions. This work was supported by the Razi University of Kermanshah, Iran and also Institute for studies in Theoretical Physics and Mathematics (IPM) of, Iran.

\end{document}